\documentclass[lettersize,transaction]{IEEEtran}

\IEEEoverridecommandlockouts
\usepackage{cite}
\usepackage{amsmath,amssymb,amsfonts}
\usepackage{algorithmic}
\usepackage{graphicx}
\usepackage{textcomp}
\usepackage{xcolor}
\usepackage{epstopdf} 
\usepackage{epsfig}
\usepackage[super]{nth}
\usepackage{subcaption}

\def\BibTeX{{\rm B\kern-.05em{\sc i\kern-.025em b}\kern-.08em
    T\kern-.1667em\lower.7ex\hbox{E}\kern-.125emX}}
\begin{document}

\title{Self-Supervised Multimodal Fusion Transformer for Passive Activity Recognition
}

\author{\IEEEauthorblockN{Armand K. Koupai,
Mohammud J. Bocus, Raul Santos-Rodriguez, Robert J. Piechocki, Ryan McConville}	\\
	\IEEEauthorblockA{School of Computer Science, Electrical and Electronic Engineering, and Engineering Maths, \\ University of Bristol, UK.}\\
	\{uw20504, junaid.bocus, enrsr, eerjp,ryan.mcconville\}@bristol.ac.uk.
}

\maketitle
\begin{abstract}
The pervasiveness of Wi-Fi signals provides significant opportunities for human sensing and activity recognition in fields such as healthcare. The sensors most commonly used for passive Wi-Fi sensing are based on passive Wi-Fi radar (PWR) and channel state information (CSI) data, however current systems do not effectively exploit the information acquired through multiple sensors to recognise the different activities. In this paper, we explore new properties of the Transformer architecture for multimodal sensor fusion. We study different signal processing techniques to extract multiple image-based features from PWR and CSI data such as spectrograms, scalograms and Markov transition field (MTF). We first propose the Fusion Transformer, an attention-based model for multimodal and multi-sensor fusion. Experimental results show that our Fusion Transformer approach can achieve competitive results compared to a ResNet architecture but with much fewer resources. To further improve our model, we propose a simple and effective framework for multimodal and multi-sensor self-supervised learning (SSL). The self-supervised Fusion Transformer outperforms the baselines, achieving a F1-score of 95.9\%. Finally, we show how this approach significantly outperforms the others when trained with as little as 1\% (2 minutes) of labelled training data to 20\% (40 minutes) of labelled training data.
\end{abstract}


\section{Introduction}
In recent years, there has been growing research interest in healthcare applications to diagnose and prevent mental and physical diseases, often within the home, and often with the objective to relieve the burden on healthcare services. Many systems have been developed to collect and provide information about a person's health condition in this way \cite{vesta}.  A wide array of sensors have been deployed, from wearables, to cameras, to more recently passive sensing systems using radio frequency (RF) signals. Sensors such as Wi-Fi are particularly promising for in-home healthcare applications, as they 1) perform sensing passively, 2) avoid any discomfort for the user (as no sensors need to be worn), 3) are ubiquitous, 4) and they are more privacy-friendly than alternatives such as cameras. Wi-Fi based sensing systems have been studied for  tasks such as language recognition \cite{languagerecognition} and fall detection \cite{falldetection}. These systems can also be used for human activity recognition (HAR) \cite{wifidata1,wifidata2} as human activities cause changes in the wireless signal transmitted by the passive WiFi sensors in terms of frequency shifts, multipath propagation and signal attenuation \cite{wificsi}. 
Two Wi-Fi sensors are commonly used in HAR, namely passive Wi-Fi radar (PWR) and channel state information (CSI). CSI represents how a wireless signal propagates from the transmitter to its receiver at particular carrier frequencies along multiple paths. The CSI data, which can be extracted from specific network interface cards (NICs) such as Intel 5300 \cite{csi_tool} or Atheros \cite{Xie}, can be viewed as a 3D time-series matrix of complex values representing both the amplitude attenuation and the phase shift of multiple propagation paths. It captures how wireless signals travel through surrounding objects or humans in time, frequency and spatial domains. 
Despite that both CSI and PWR sensors use the same signal source and have a similar function, PWR works differently. A PWR system correlates the transmitted signal from a WiFi access point and the reflected signal from the surveillance area and calculate the distance between the antenna and the object or human \cite{csipwr}.

Research in radio-based human sensing and activity recognition has moved towards deep learning, principally because deep learning models can learn complex representations from raw and noisy data. However, most passive HAR deep learning based systems are uni-modal, i.e., they use information from only one type of sensor. These systems usually use two different architectures: convolutional neural networks (CNN) which have been principally used with RF sensors' raw data transformed into image-like representation such as spectrograms \cite{cnn}, or recurrent neural networks (RNN) which work directly on the raw Wi-Fi data \cite{rnn}. In this work, we study and propose to use multiple synchronised sensors, or views, in an indoor environment to improve the performance of a passive HAR system. More specifically, we propose to build a system that collects raw data from synchronized sensors, and after some signal processing, all modalities can be fused effectively; in this case using Transformers.

Recent work has demonstrated that the Vision Transformer (ViT) \cite{dosovitskiy2020vit} architecture is capable of competitive or superior performance on image classification tasks at a large scale. Instead of convolutions, it uses a self-attention mechanism to aggregate information across locations. Thus, we investigate the potential of the ViT for sensor and feature fusion. For this purpose, we need to address the possible challenges: firstly, how the self-attention mechanism present in the ViT can be used for sensor feature fusion? Secondly, does ViT benefit from sensor fusion and lead to better predictions for HAR. In this paper, we study these questions and compare our findings with a traditional ResNet model. The main contributions of this work are the following:
\begin{itemize}
\item We propose a model, the Sensor Fusion Vision Transformer (SF-ViT), based on the Vision Transformer (ViT) architecture \cite{dosovitskiy2020vit}, that can fuse multiple image features and views.
\item We extend it to a more general framework which we called the Fusion Transformer, that can effectively fuse multiple features from different types of sensors and we assess the effectiveness of our transformer-based model for multi-modal and multi-sensor fusion.
\item We evaluate the effectiveness of the Fusion Transformer on a human activity recognition (HAR) dataset (collected using Wi-Fi sensors) in a purely supervised fashion, and compare its performance against ResNet.
\item We also propose a new method for multi-modal and multi-sensor self-supervised learning (SSL) that outperforms the baselines using multiple image features and views for passive HAR.
\end{itemize}

This paper is organised as follows: Related works on multimodal sensor fusion are presented in Section II. 
The methodology and system design of our multimodal sensor fusion transformer models are described in Section III, including details on the signal processing of WiFi-based signals.
Section IV provides detailed information on the experimental setup. 
Section V presents the results obtained using our fully supervised Fusion Transformer model on a human activity recognition dataset.
Section VI describes our self-supervised multimodal sensor fusion transformer architecture, along with details on the experiment setup and results.
Finally, conclusions are drawn in Section VII.

\section{Related Work}
Most works on multi-modal or multi-sensor fusion for human action recognition using RF, inertial and/or vision sensors, have considered either decision-level fusion or feature-level fusion. 
For example, the authors of \cite{wivi} perform multimodal fusion at the decision level to combine the advantages of Wi-Fi and vision-based sensors using a hybrid deep neural network (DNN) model to achieve a 97.5\% cross-validation accuracy on average for 3 activities: sitting, standing and walking.
The model essentially consists of a Wi-Fi sensing module (CNN architecture) and a vision sensing module (based on the convolutional 3D model) for processing Wi-Fi and video frames for unimodal inference, followed by a multimodal fusion module. 
Multimodal fusion is performed at the decision level (after both the Wi-Fi and vision modules have made a classification) because this framework is stated to be more flexible and robust to unimodal failure compared to feature level fusion. 
The authors of \cite{gimme} present a method for HAR, which leverages four sensor modalities, namely, skeleton sequences, inertial and motion capture measurements and Wi-Fi fingerprints.
The fusion of signals is formulated as a matrix concatenation. The individual signals of different sensor modalities are transformed and represented as an image. The resulting images are then fed to a 2D CNN (EfficientNet B2) for classification. The authors evaluated their approach on four different datasets; the NTU RGB+D 120 dataset for skeleton data, the UTD-MHAD dataset for skeleton and inertial data, the ARIL dataset for Wi-Fi data and the Simitate dataset for motion capture data. Good experimental results were achieved across the different sensor modalities.
The authors of \cite{wiwehar} proposed a multimodal HAR system that leverages Wi-Fi and wearable sensor modalities to jointly infer human activities. They collected CSI data from a standard Wi-Fi NIC, alongside the user's local body movements via a wearable inertial measurement unit (IMU) consisting of an accelerometer, gyroscope, and magnetometer sensors. They calculated the time-variant mean Doppler shift (MDS) from the processed CSI data and magnitude from the inertial data for each sensor of the IMU. 
Then, various time and frequency domain features were separately extracted from the magnitude data and the MDS. The authors applied a feature-level fusion method which sequentially concatenates feature vectors that belong to the same activity sample. Finally, supervised machine learning techniques were used to classify four activities, such as walking, falling, sitting, and picking up an object from the floor.
The authors of \cite{csipwr} conducted a comprehensive study on the comparison of two RF sensing devices for the purpose of HAR, 
namely, CSI and PWR systems. Two pipelines were proposed for filtering and processing the raw signals from the two sensors into Doppler spectrograms, which were then used to train a simple supervised CNN to evaluate the HAR performance. They considered the combined activity data from 3 different layouts. In the first layout, the transmitter and receiver were facing each other (in a line-of-sight configuration) while in the second layout, the transmitter and receiver were at 90$^\circ$ to each other. Finally, in the third layout, the transmitter and receiver were co-located (placed next to each other).
The CSI system achieved an overall accuracy of 67,3\% while the PWR system had an accuracy of 66,7\%.
Although this work presents a simple system which combines CSI and PWR spectrograms by merging probabilities from two networks (decision-level fusion), current state-of-the-art models are not specifically designed for the fusion of multiple passive Wi-Fi devices.

While CNN architecture was the de-facto standard for computer vision tasks, gradually, ViT showed very promising results when pre-trained on large amounts of data and then fine-tuned to mid-sized or small-sized image recognition benchmarks while requiring fewer computational resources for training \cite{dino,beit}. However, most of ViT models have been trained on natural images of very large size, together with pre-training and very strong data augmentation techniques. A similar work which also trained a ViT with spectrograms is the audio spectrogram transformer (AST) \cite{ast}, which presents a new method for audio classification with a ViT using spectrogram data. Recent works showed that ViTs could outperform ResNets without pre-training or strong data augmentations \cite{resnetvit}, notably by using sharpness-aware minimisation technique \cite{sam}, which simultaneously minimises the loss value and loss sharpness by seeking parameters that lie in neighbourhoods and having uniformly low loss. However, this technique requires the computation of two forward-backward propagations to estimate the `sharpness-aware' gradient, and thus, leads to an increased training time.

In this paper, we evaluate the performance of our Transformer-based sensor fusion model for HAR using image data generated from multiple sensors. We evaluate its potential for sensor fusion and propose a method for multi-modal and multi-sensor self-supervised learning (SSL).
\begin{figure}[!htp]
    \centering
    \includegraphics[width=88mm,scale=0.5]{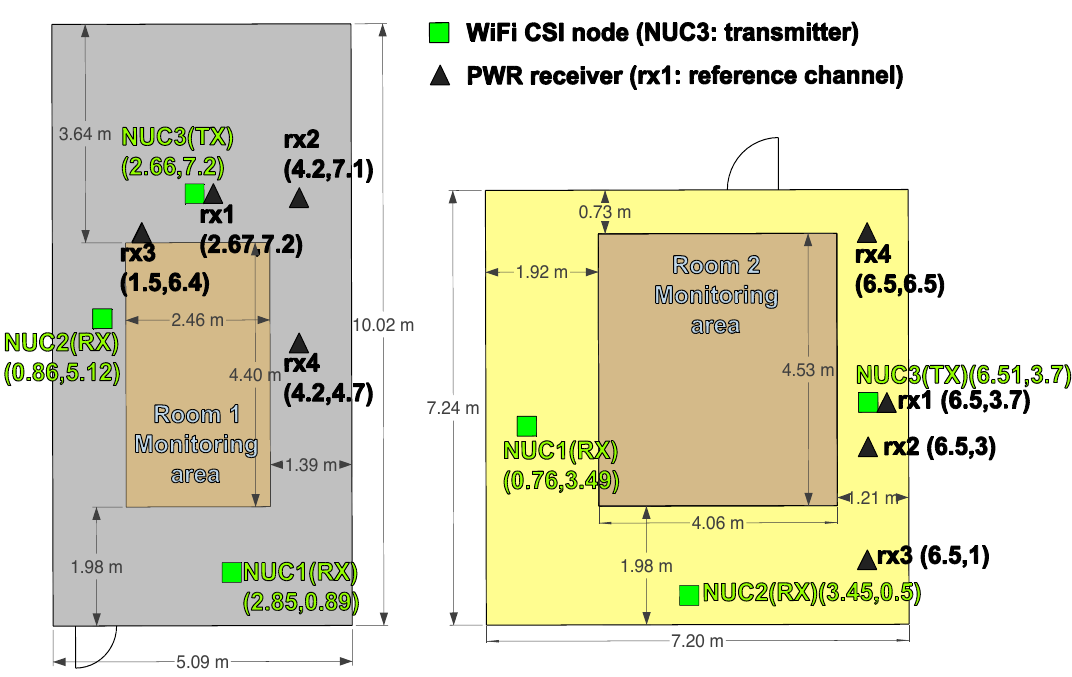}
    \caption{The CSI system and PWR system deployment \cite{operanet}. The CSI system consisted of 2 receivers (denoted as NUC1 and NUC2) while the PWR system consisted of 3 receivers (denoted as rx2, rx3, and rx4). For more details on the dataset, the interested reader is referred to \cite{operanet}.}
    \label{fig:setup}
\end{figure}

\begin{figure*}[ht]
  \begin{subfigure}[a]{\textwidth}
    \includegraphics[width=\textwidth]{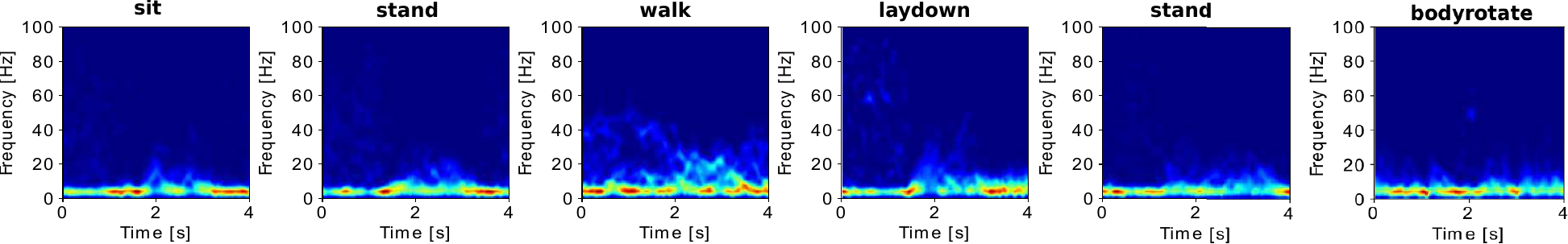}
    \caption{Amplitude CSI spectrograms from view 1 (NUC1) for each activity.}
    \label{fig1:fa}
  \end{subfigure}
  \vfill
  \begin{subfigure}[b]{\textwidth}
    \includegraphics[width=\textwidth]{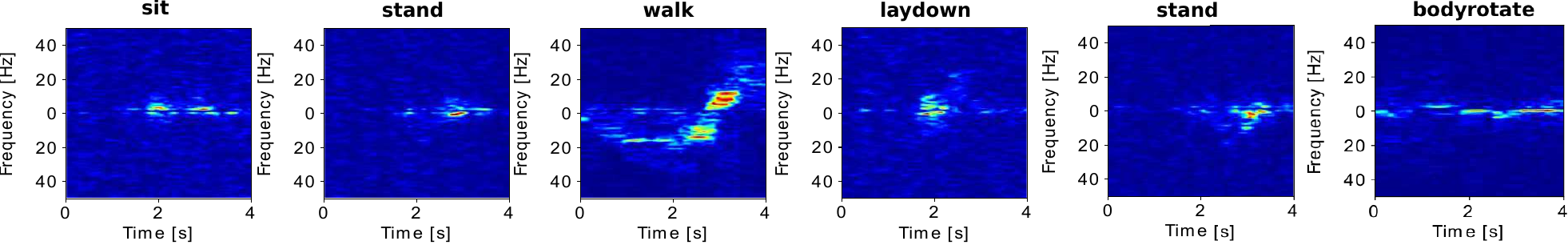}
    \caption{PWR spectrograms from first channel for each activity.}
    \label{fig1:fb}
  \end{subfigure}
  \caption{Visualisation of PWR and CSI spectrograms for each activity.}
  \label{fig1:f}
\end{figure*}

\section{Methodology and System Design}
\subsection{Signal Processing of RF Sensors}
Inspired by other work in this area \cite{csipwr}, which has explored two pipelines for extracting image features from RF sensors using signal processing techniques, we apply the same principles. In this work, we use the OPERAnet dataset \cite{operanet}, which includes publicly available data from both CSI and PWR systems. 
The dataset was collected with the intention to evaluate HAR and localisation techniques with measurements obtained from synchronized RF devices and vision-based sensors. 
The experimental setup established to collect both CSI and PWR data is shown in Fig. \ref{fig:setup}.
Fig. \ref{fig1:fa} and \ref{fig1:fb} show some examples of the generated spectrograms with these two pipelines, for each of the six activities, namely, sitting down on a chair ('\textit{sit}'), standing from chair ('\textit{stand}'), laying down on the floor ('\textit{laydown}'), standing from floor ('\textit{standff}'), body rotation ('\textit{bodyrotate}'), and walking ('\textit{walk}'). The pipelines are as follows:
\begin{itemize}
    \item In Fig. \ref{fig1:fa}: we denoise the CSI signal using discrete wavelet transform (DWT) and median filtering, then reduce the dimensionality using principal component analysis (PCA) and generate a spectrogram using short time Fourier transform (STFT).
    \item In Fig. \ref{fig1:fb}: we apply cross ambiguity function to the raw PWR data, use the CLEAN algorithm and constant false alarm rate (CFAR) for direct signal cancellation and noise reduction, generating as output a Doppler spectrogram \cite{csipwr}.
\end{itemize}

These two pipelines are necessary to extract informative data from CSI and PWR sensors. The raw CSI data is very noisy in nature, and thus the DWT technique helps to filter out high frequency components and remove noises, while preserving most of the information and avoiding the distortion of the signal \cite{translation_resilient}. Afterwards, we perform median filtering to remove any undesired transients in the CSI mesurements which have not been performed by human motion. Despite that the CSI data is highly informative, it consists of a lot of complex values per second, depending on the number of transmit and receive antennas, orthogonal frequency-division multiplexing (OFDM) subcarriers and packet rate (for example, the Intel 5300 chipset captures complex CSI data over 3 transmit antennas, 3 receive antennas and 30 subcarriers).
Therefore, we use PCA to reduce the computational complexity of such data,
while preserving as much information as possible. Finally, we convert the PCA signal into spectrograms using STFT \cite{csipwr}. 

For the PWR signal, we first apply the cross ambiguity function (CAF) to extract target range and Doppler information. However, we also capture an interference source which is the strong direct signal emitted from the Wi-Fi access point and which is captured by the PWR surveillance channel. Thus, to remove this signal, we employ the CLEAN algorithm \cite{clean}. The last step consists of reducing the noise on the CAF surface. We use CFAR to estimate the background noise and apply it to the CAF surface. PWR's Doppler spectrogram is generated by selecting the maximum Doppler pulse from each Doppler bin within the CAF surface \cite{csipwr}.

It should be noted that all the devices were synchronised to the same network time protocol (NTP) server and were labelled in sync. Thus, the raw data could be segmented as per the ground truth activity labels and processed accordingly.

\begin{figure*}[!htp]
\centering
\captionsetup{justification=centering,margin=2cm}
  \begin{subfigure}[b]{.15\textwidth}
    \includegraphics[width=\textwidth]{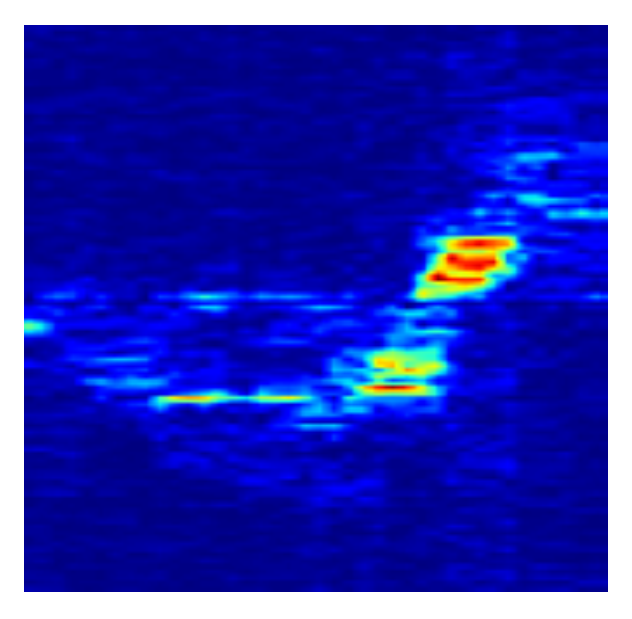}
    \caption{PWR channel 1  \\}
  \end{subfigure}
  \hfill
  \begin{subfigure}[b]{.15\textwidth}
    \includegraphics[width=\textwidth]{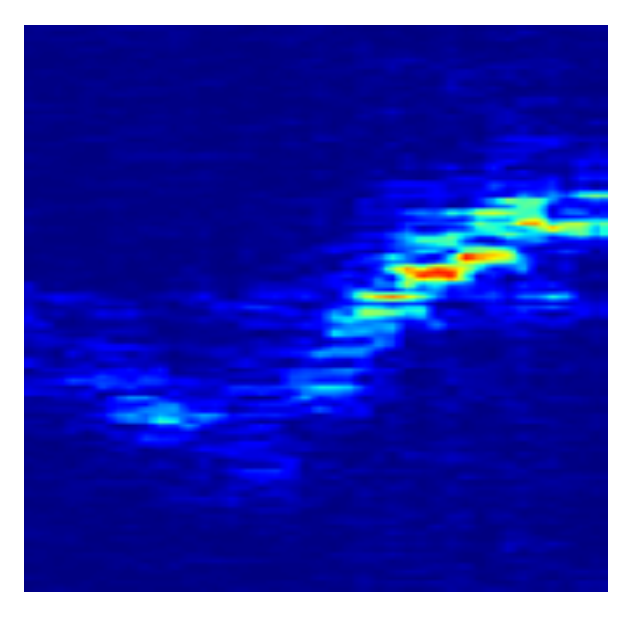}
    \caption{PWR channel 2}
  \end{subfigure}
  \hfill
  \begin{subfigure}[b]{.15\textwidth}
    \includegraphics[width=\textwidth]{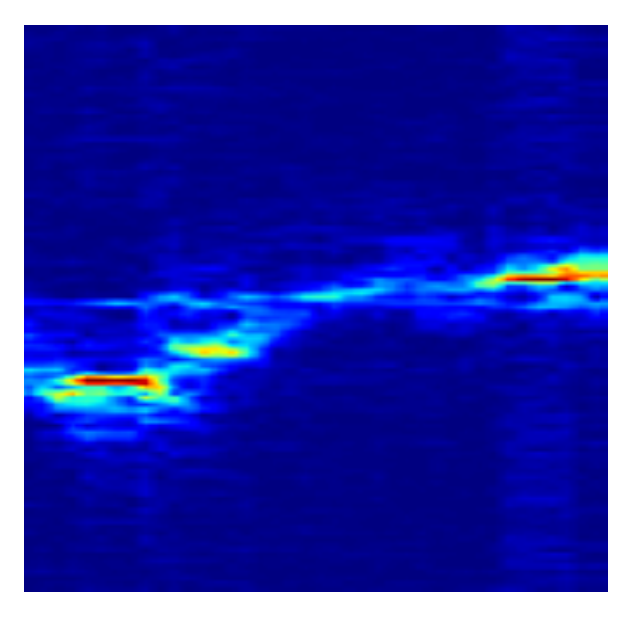}
    \caption{PWR channel 3}
  \end{subfigure}
  \hfill
  \begin{subfigure}[b]{.15\textwidth}
    \includegraphics[width=\textwidth]{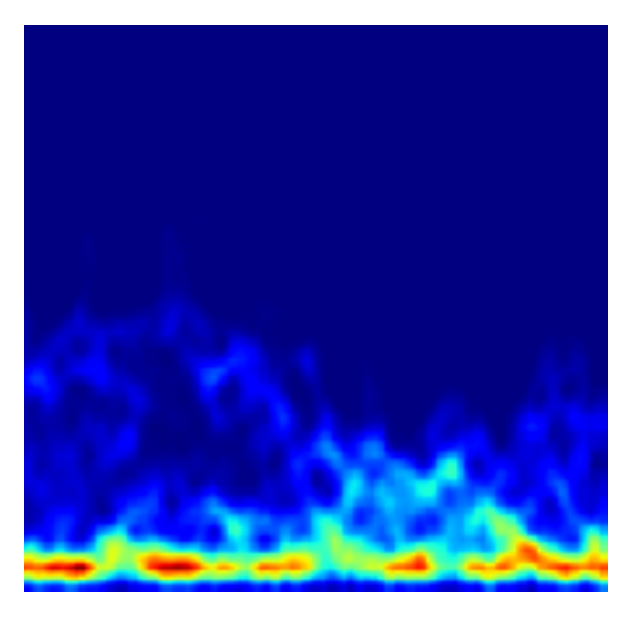}
    \caption{Amp. spec. N1}
  \end{subfigure}
  \hfill
  \begin{subfigure}[b]{.15\textwidth}
    \includegraphics[width=\textwidth]{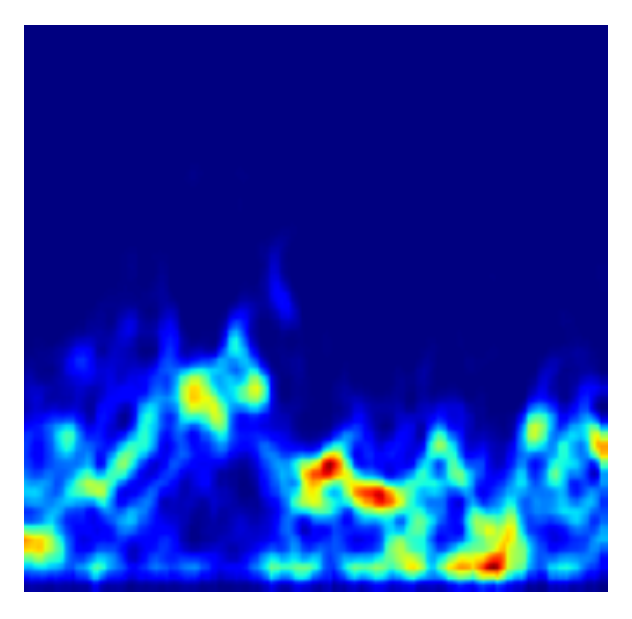}
    \caption{Amp. spec. N2}
  \end{subfigure}
  \vfill
  \begin{subfigure}[b]{.15\textwidth}
    \includegraphics[width=\textwidth]{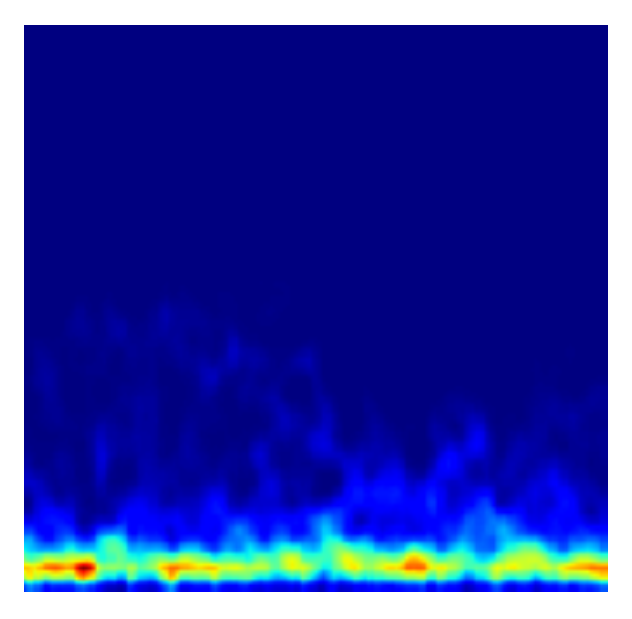}
    \caption{Ph. diff. N1}
  \end{subfigure}
  \hfill
  \begin{subfigure}[b]{.15\textwidth}
    \includegraphics[width=\textwidth]{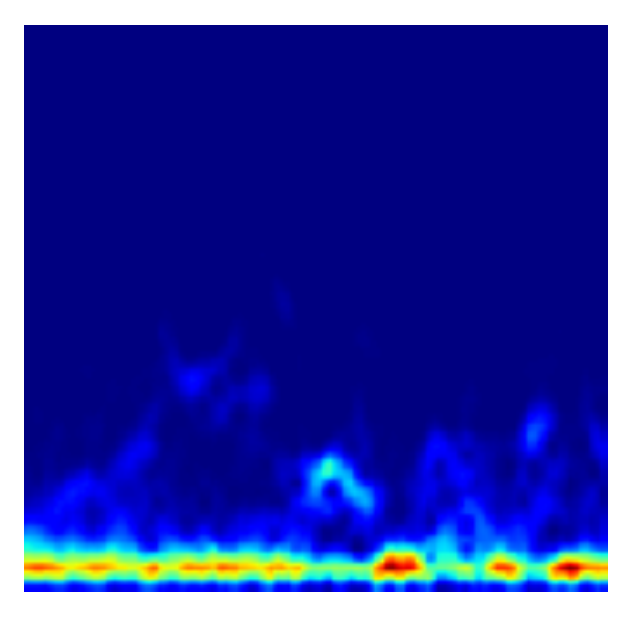}
    \caption{Ph. diff. N2}
  \end{subfigure}
  \hfill
  \begin{subfigure}[b]{.15\textwidth}
    \includegraphics[width=\textwidth]{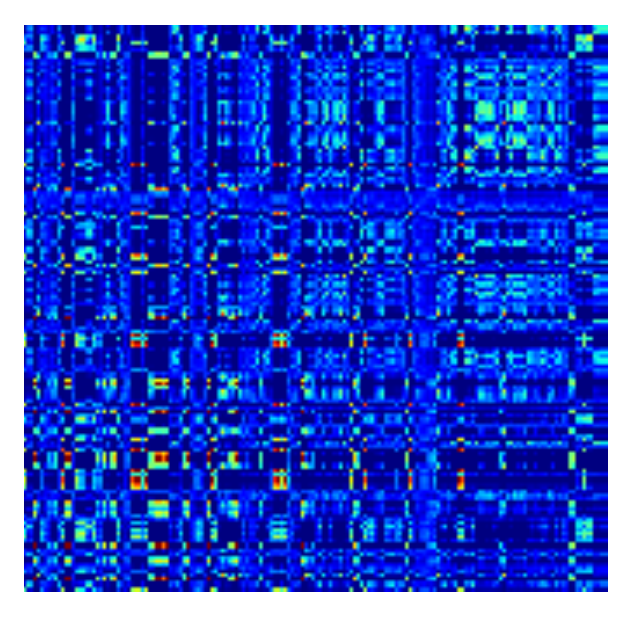}
    \caption{MTF ph. diff. N1}
  \end{subfigure}
  \hfill
  \begin{subfigure}[b]{.15\textwidth}
    \includegraphics[width=\textwidth]{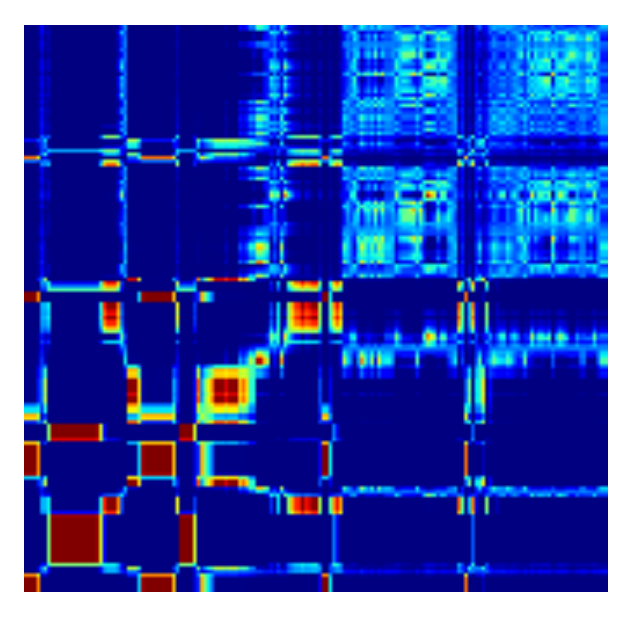}
    \caption{MTF ph. diff. N2}
  \end{subfigure}
  \hfill
  \begin{subfigure}[b]{.15\textwidth}
    \includegraphics[width=\textwidth]{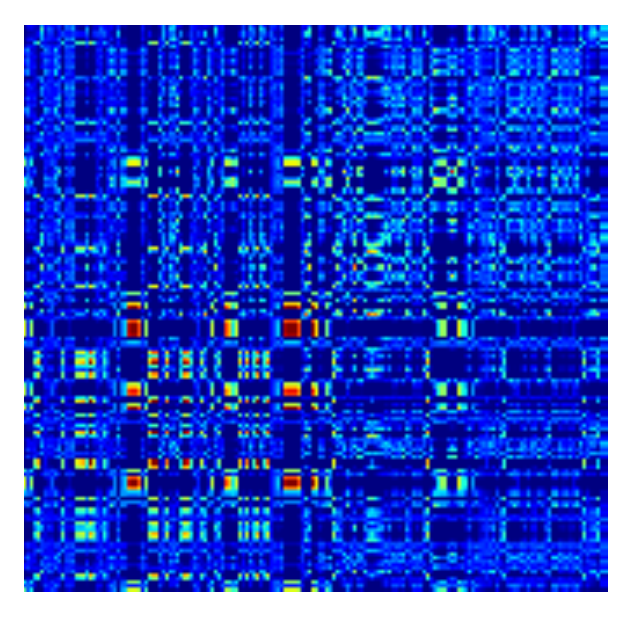}
    \caption{MTF amp. N1}
  \end{subfigure}
  \vfill
  \begin{subfigure}[b]{.15\textwidth}
    \includegraphics[width=\textwidth]{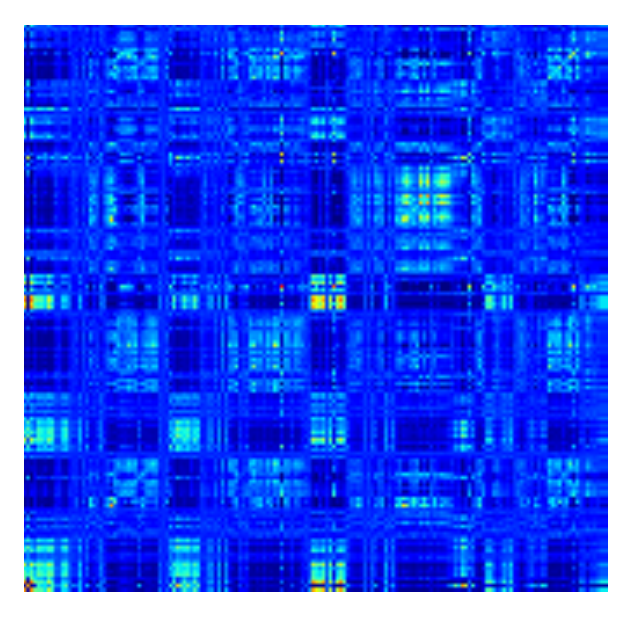}
    \caption{MTF amp. N2}
  \end{subfigure}
  \hfill
  \begin{subfigure}[b]{.15\textwidth}
    \includegraphics[width=\textwidth]{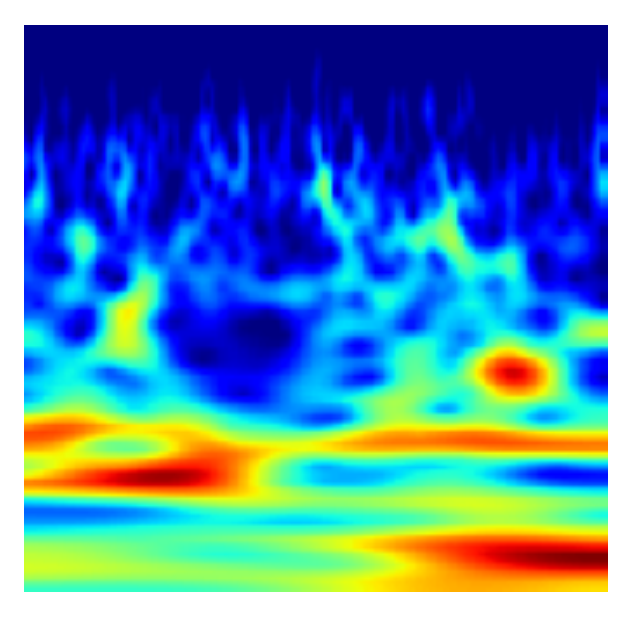}
    \caption{Amp. scal. N1}
  \end{subfigure}
  \hfill
  \begin{subfigure}[b]{.15\textwidth}
    \includegraphics[width=\textwidth]{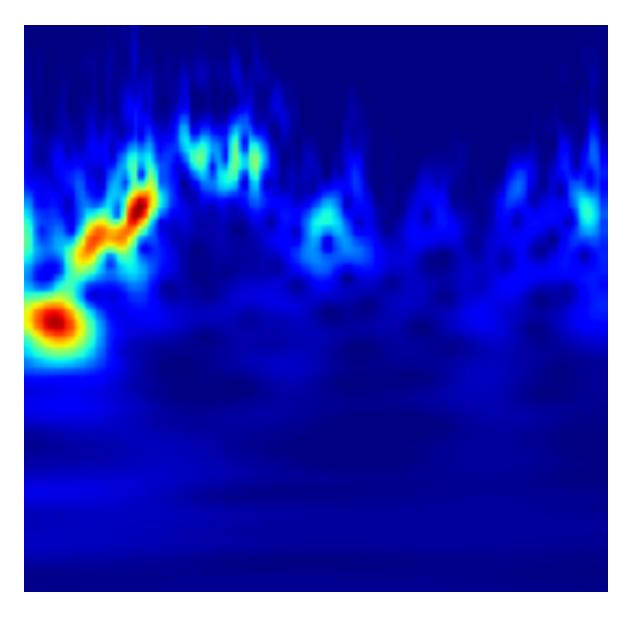}
    \caption{Amp. scal. N2}
  \end{subfigure}
  \hfill
  \begin{subfigure}[b]{.15\textwidth}
    \includegraphics[width=\textwidth]{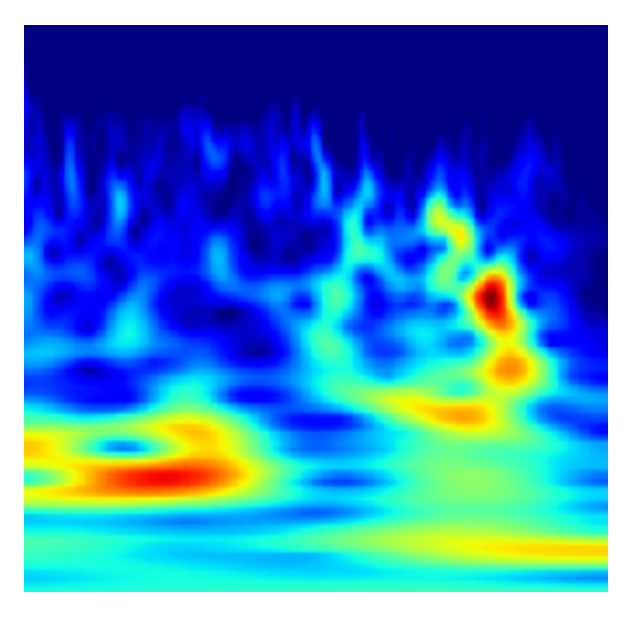}
    \caption{Ph. diff. scal. N1}
  \end{subfigure}
  \hfill
  \begin{subfigure}[b]{.15\textwidth}
    \includegraphics[width=\textwidth]{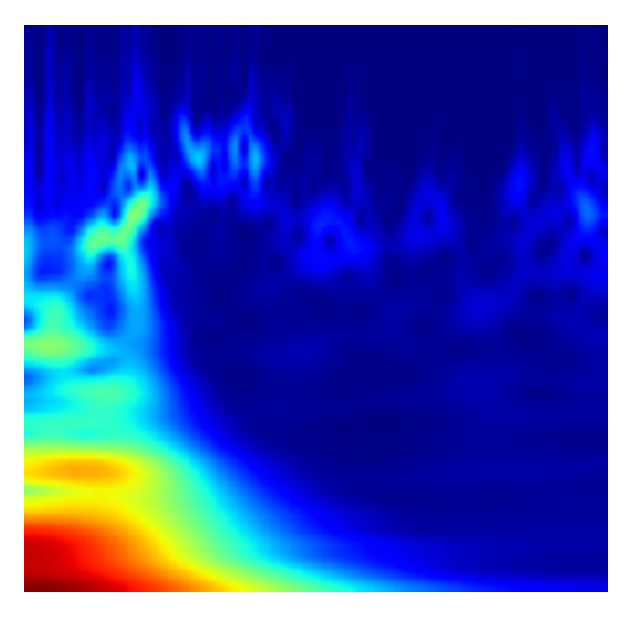}
    \caption{Ph. diff. scal. N2}
  \end{subfigure}
  \caption{Fifteen image features extracted via signal processing techniques representing a person walking in the monitoring area for a duration of four seconds.}
  \label{fig:features}
\end{figure*}

Using the CSI data, we also generate
other features such as scalograms and Markov transition fields (MTF). Each feature captures particular information about the activity. Given all of these different features, we aim to build a network that can fuse all these images together effectively to improve the overall system performance. In this work, we have extracted 15 different features (see Fig. \ref{fig:features}):
\begin{itemize}
    \item PWR spectrogram data collected from the three  receiver surveillance channels, rx2, rx3, and rx4 in Fig. \ref{fig:setup}.   
    (denoted as `\textit{PWR channel 1}', `\textit{PWR channel 2}', `\textit{PWR channel 3}', respectively, in Fig. \ref{fig:features});
    \item Spectrograms generated using STFT on amplitude CSI data from the two receivers, NUC1 (`\textit{Amp. spec. N1}') and NUC2 (`\textit{Amp. spec. N2}');
    \item Spectrograms generated using STFT on phase difference CSI data for each of the two receivers (`\textit{Ph. diff. N1}', `\textit{Ph. diff. N2}');
    \item Markov transition field (MTF) \cite{mtf} features generated from phase difference CSI data acquired from two receivers (`\textit{MTF ph. diff. N1}', `\textit{MTF ph. diff. N2}');
    \item  MTF features generated from amplitude CSI data acquired from two receivers (`\textit{MTF amp. N1}', `\textit{MTF amp. N2}');
    \item  Scalograms 
    generated by applying continuous wavelet transform (CWT) on the amplitude CSI data from NUC1 (`\textit{Amp. scal. N1}') and NUC2 (`\textit{Amp. scal. N2}') receivers;
    \item Scalograms generated using CWT on the phase difference CSI data from the two CSI receivers (`\textit{Ph. diff. scal. N1}', `\textit{Ph. diff. scal. N2}').
\end{itemize}
Each channel and receiver can be seen as another view of the human activity performed in the room. Previously, we presented the spectrograms of CSI and PWR data, which give a visual representation of the spectrum of frequencies of a signal varying through time. 
Spectrograms are generated through STFT by applying a sliding window to obtain equally-sized segments of the signal and then FFT is performed on the samples in each segment, which converts the signal from the time domain to the frequency domain. 
Similar to STFT, the scalogram is a time-frequency representation of a signal and it is obtained from the absolute value of the CWT of a signal.
Finally, we also introduced another type of representation called the Markov transition field (MTF), which is an image generated from time series data, representing a field of transition probabilities for a discretized time series.

\begin{figure*}[!htp]
    \includegraphics[width=\textwidth]{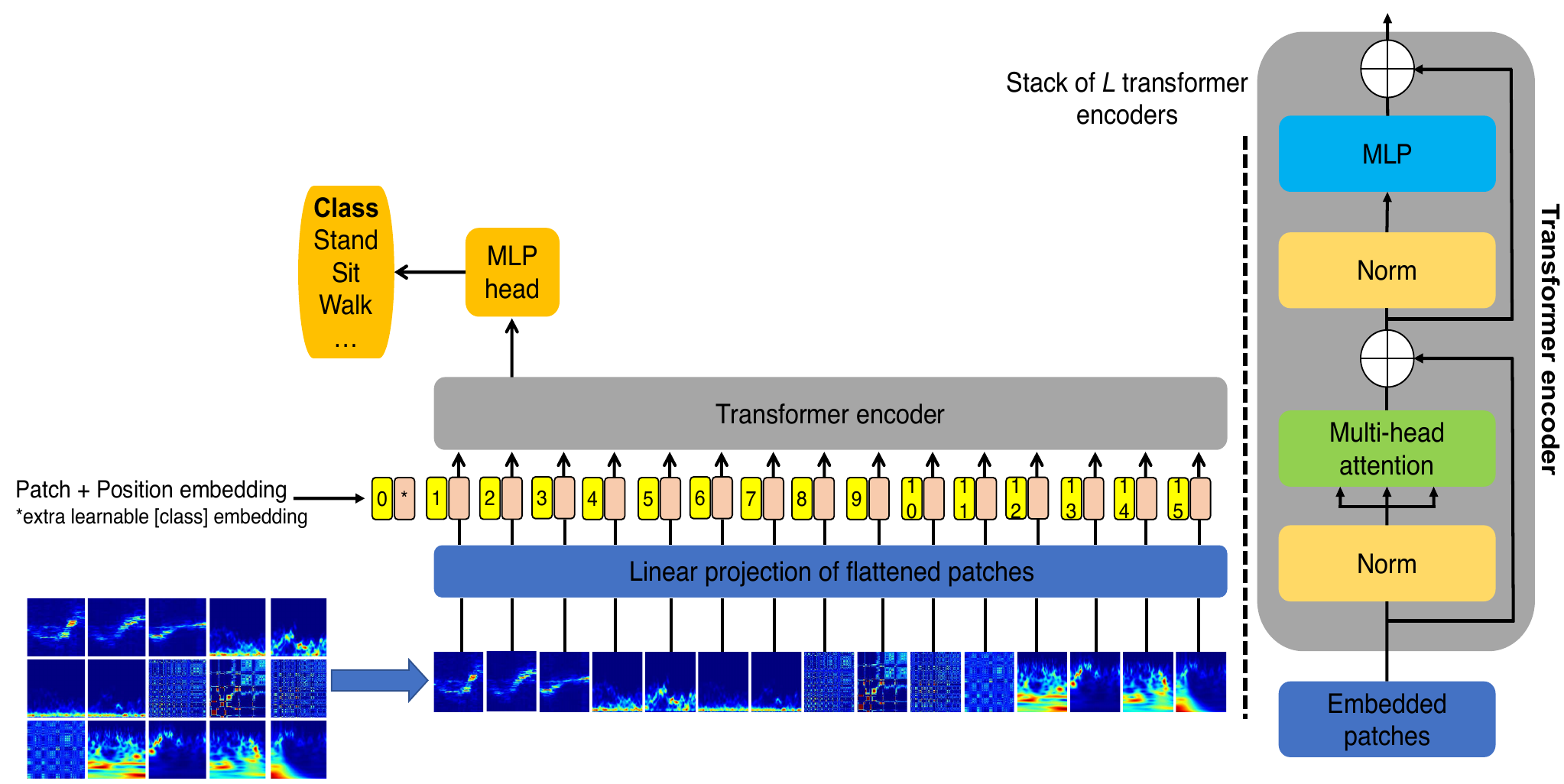}
    \caption{Sensor-Fusion Vision Transformer (SF-ViT) model overview. We split our concatenated image into patches of fixed size, 224 $\times$ 224, where each patch corresponds to one of the image features. We linearly embed each of them, add position embeddings, and feed the output embedding to the transformer encoder. In order to perform classification, we add an extra learnable `classification token'. The network model is inspired from the original ViT architecture \cite{dosovitskiy2020vit}.}
    \label{fig:f3}
\end{figure*}
\subsection{Multimodal Sensor Fusion Transformer}
\subsubsection{A first approach: Sensor Fusion Vision Transformer}
We will first present the Sensor-Fusion Vision Transformer (SF-ViT), which uses a similar architecture to the conventional Vision Transformer (ViT). Nevertheless, in most applications where ViT is used, the model is trained with `natural' images of size 224 $\times$ 224 $\times$ 3 (height, width, channels) that are divided into small patches of size 16 $\times$ 16 or 32 $\times$ 32. Here instead, we concatenate all image features and obtain an image of size 224 $\times$ ( 224 $\times$ $N$ ) $\times$ 1 where $N$ is the number of different image features concatenated. Instead of dividing our image into small patches of size 16 $\times$ 16, we patch the image so that each patch represents a different image-based feature. Fig. \ref{fig:f3} illustrates an overview of the SF-ViT, where the shape of the input image has been changed for convenience. 

The SF-ViT trains a transformer to recognise human activities by assigning a high attention weight to relevant features (i.e., our patches of different features), and a low attention weight to less pertinent image features. The SF-ViT's inspiration is that the more unique the image features that are used with the ViT are, the more effective is the model for recognising human activities, as each image feature will represent or capture different information about the activity, and thus combining them effectively should lead to better performance.

\subsubsection{The Fusion Transformer}
\begin{figure*}[!htp]
    \includegraphics[width=\textwidth]{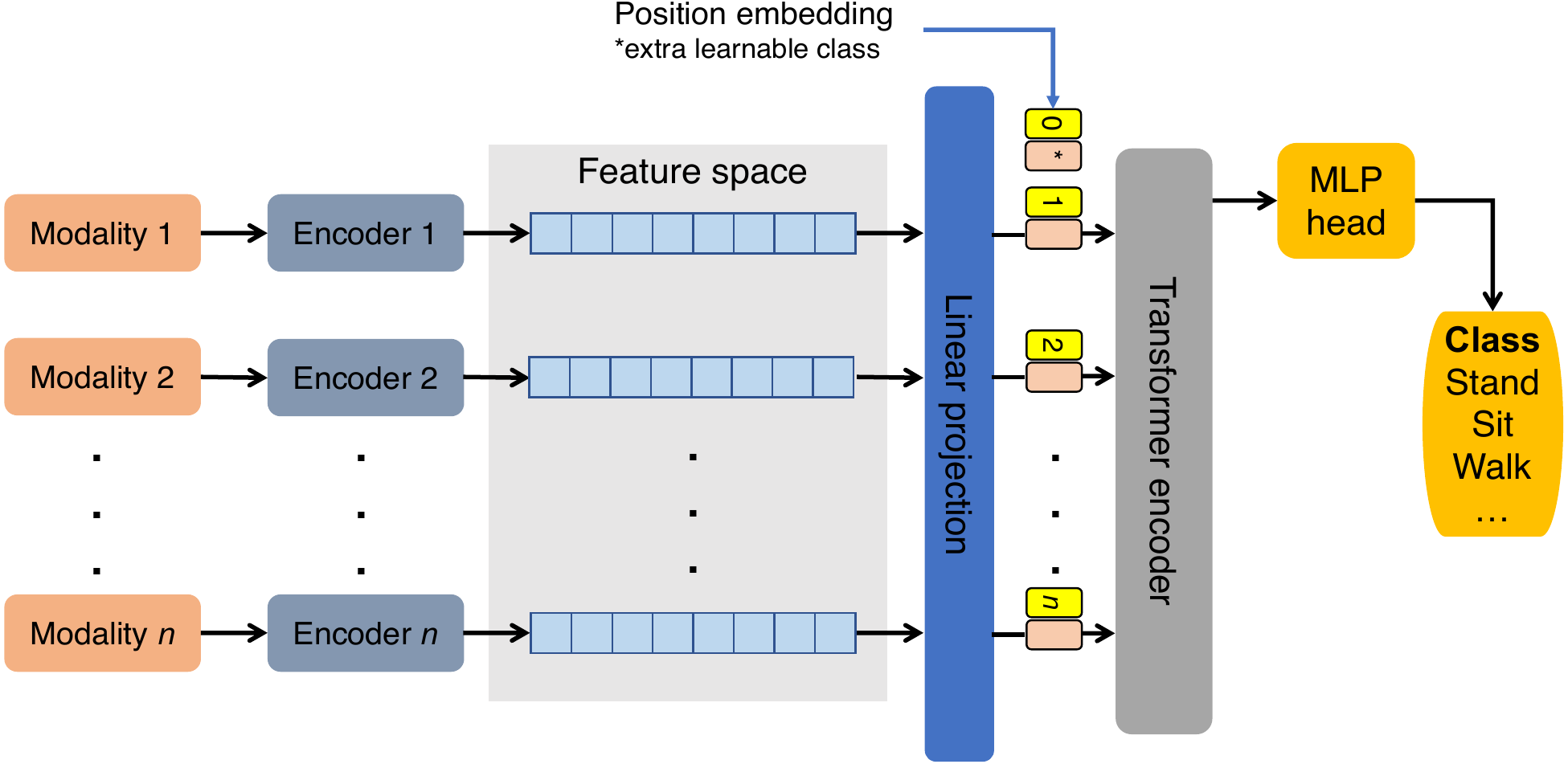}
    \captionsetup{font={normalsize}}
    \caption{The Fusion Transformer model overview. Each modality is encoded into a new feature space and then linearly embedded. We add position embeddings and feed the output embedding to the transformer encoder. We add an extra learnable classification token.}
    \label{fig:mmtrans}
\end{figure*}

One of the potential issues with this approach is that we do a linear projection of patches of size 224 $\times$ 224 $\times$ 1 into a feature space of size 512, which is computationally expensive and results in a very large number of trainable parameters and potential over-fitting, as each input pixel is connected to the linear layer. To remedy this, we instead first encode each image-based feature using a CNN encoder, which transforms the raw image feature of size 224 $\times$ 224 $\times$ 1 into an image of size 16 $\times$ 16 $\times$ 64. This new architecture can be considered as a multi-modal model, where each modality is first passed into an encoder that transforms the raw modality into a smaller feature space. The corresponding model architecture, which we call Fusion Transformer, is presented in Fig. \ref{fig:mmtrans}.

\begin{table*}[ht]
\centering
\caption{ViT parameters.}
\begin{tabular}{|c|c|c|c|c|c|c|c|}
\hline
image size & patch size & channels & emb. dim. & depth & qkv bias & drop out & MLP ratio \\ \hline
224, $N$ $\times$ 224 & 224, 224 & 1 & 512 & 3 & False & 0.1 & 1.0 \\ \hline
\end{tabular}
\label{tab:t1}
\end{table*}
\section{Experimental setup}
We evaluate the capabilities of our Fusion Transformer on human activity recognition and compare its performance to ResNet and show that the Fusion Transformer is successful in achieving competitive results while requiring less computational resources than ResNet. In this section, the experimental setup used throughout the findings of the paper is presented. The system was developed in PyTorch and all models have been trained on a single GPU (Nvidia 2080Ti).
\subsection{Dataset and Metrics}
As mentioned previously, for our experimentation, we will use the OPERAnet dataset \cite{operanet}, which includes publicly available data from both CSI and PWR systems. 
The RF sensors captured the changes in the wireless signals while six daily activities were being performed by six participants, namely, sitting down on a chair ('\textit{sit}'), standing from chair ('\textit{stand}'), laying down on the floor ('\textit{laydown}'), standing from floor ('\textit{standff}'), body rotation ('\textit{bodyrotate}'), and walking ('\textit{walk}').
Applying the signal processing pipelines described earlier, led to a dataset composed of 2,897 data samples (non-overlapping windows each representing 4 seconds of an activity) for the six activities. Worth noting however, as is the case in reality, the distribution of the different activities a human engages in is highly imbalanced. In this case,  we have an imbalanced dataset where the two most represented classes are body rotating and walking, representing respectively 30\% and 33\% of the total observations. The two classes which are less represented are `standing from floor' and `laying down', each representing 7\% of the dataset. For training and validation purposes, we randomly split the dataset into a train set and a  validation set, respectively composed of 80\% and 20\% of the total dataset samples. For this dataset, we use the accuracy and macro F1-score as our main metrics.
\begin{table*}[!htp]
\centering
\caption{Performance comparison of the Fusion Transformer with ResNet.}
%\resizebox{\textwidth}{!}{
\begin{tabular}
%{c|c|c|c|c|c|c|c|c}
{p{0.33\linewidth} | p{0.055\linewidth} | p{0.055\linewidth} | p{0.055\linewidth}| p{0.055\linewidth}| p{0.055\linewidth}| p{0.055\linewidth}| p{0.055\linewidth}| p{0.055\linewidth} }
 & \multicolumn{2}{c}{\textbf{SF-ViT}} & \multicolumn{2}{c}{\textbf{ResNet18}} & \multicolumn{2}{c}{\textbf{ResNet34}} & \multicolumn{2}{c}{\textbf{Fusion Transformer}}  \\ \hline
& Accuracy  & F1-score & Accuracy    & F1-score      & Accuracy   & F1-score & Accuracy & F1-score
 \\ \hline
CSI amplitude spectrogram (view 1 = 1 feature)                & 80.3\%      & 71.7\% & 92.8\%      & 89.7\%             & 73.3\%     & 71.0\%              & 88.3\%     & 83.5\%  \\ \hline
CSI amplitude spectrogram (2 views = 2 features)               & 85.9\%      & 78.6\% & 93.1\%      & 90.0\%             & 35.6\%     & 43.4\%              & 92.1\%     & 87.0\%  \\ \hline
CSI ph. diff + amp. spectrograms (4 features)       & 84.3\%      & 77.4\% & 94.5\%      & 91.4\%             & 95.7\%     & 93.9\%              & 92.2\%     & 88.4\%  \\ \hline
CSI (amp. + ph. diff.) \& PWR spectrograms (7 features)          & 91.6\%      & 88.2\% & \textbf{95.0\%}      & \textbf{92.4\%}             & \textbf{96.6\%}     & \textbf{94.9\% }    & \textbf{95.9\%}     & \textbf{94.3\%} \\ \hline
CSI (amp. + ph. diff.) spectrograms + PWR spectrograms + CSI (amp. + ph. diff.) MFT (11 features)        & 91.9\%      & 88.6\% & 93.8\%      & 91.1\%             & 91.9\%     & 88.2\%              & 94.3\%     & 91.9\%  \\ \hline
All 15 image features                & \textbf{92.8\%}      & \textbf{89.5\%} & 91.0\%      & 86.5\%             & 93.3\%     & 90.4\%              & 93.6\%     & 91.1\% \\    
\end{tabular}
%}
\end{table*}
\subsection{Models}
\label{sec:s1}
In this section, we will initially focus on the Fusion Transformer. All experiments have been performed with the configuration presented in Table \ref{tab:t1}.
The width of the image is $N$ $\times$ 224, where $N$ is the number of different image features generated. We compare and train the model with different numbers of features to analyse how the model’s performance scales. 

In the Fusion Transformer shown in Fig. \ref{fig:mmtrans}, we add a CNN encoder for our images to extract more relevant features from the raw images and to reduce the size of the images. The CNN encoder is composed of 4 blocks, where each block consists of a convolution, ReLU and pooling layers. Each image feature of size $ 224 \times 224 \times 1$ is embedded in a new image representation of dimension $16 \times 16 \times 64$.

To compare the performance of the Fusion Transformer with a baseline, we also train a ResNet model to evaluate whether it achieves better performance when trained with multiple image features, by considering each feature as a new channel. We trained two models: ResNet18 and ResNet34.
\subsection{Training}
All models, including ResNet, have been trained using the AdamW optimizer, with $\beta{_1} = 0.90$ and $\beta{_2} = 0.999$, with a weight decay settled at 0.01 and a batch size of $64$. The learning rate has been initialised at 1e-4 and reduced during training using a learning step scheduler with a unitary step size and $\gamma=0.5$. The loss function used for these experiments is cross-entropy.

Despite that recent works train ViTs using pre-training or transfer learning on large datasets, we decided to train our model from scratch, to more closely study the benefit of our sensor fusion model for activity recognition.
Furthermore, when training on smaller datasets, ViT-based models have a weaker inductive bias compared to CNNs and thus leads to an increased reliance on model regularisation or data augmentation \cite{trainvit}. In the case of CSI and PWR data, using similar data augmentation techniques as those used on natural images is not possible. Thus, throughout all our experiments, we did not use data augmentation. We used a simple dropout strategy as mentioned in section \ref{sec:s1} and a weight decay equal to $0.01$.

\section{Results}
\subsection{Fully Supervised Fusion Transformer Results}
\begin{figure}[ht]
  \centering
  \begin{subfigure}[b]{0.8\columnwidth}
    \includegraphics[width=\columnwidth]{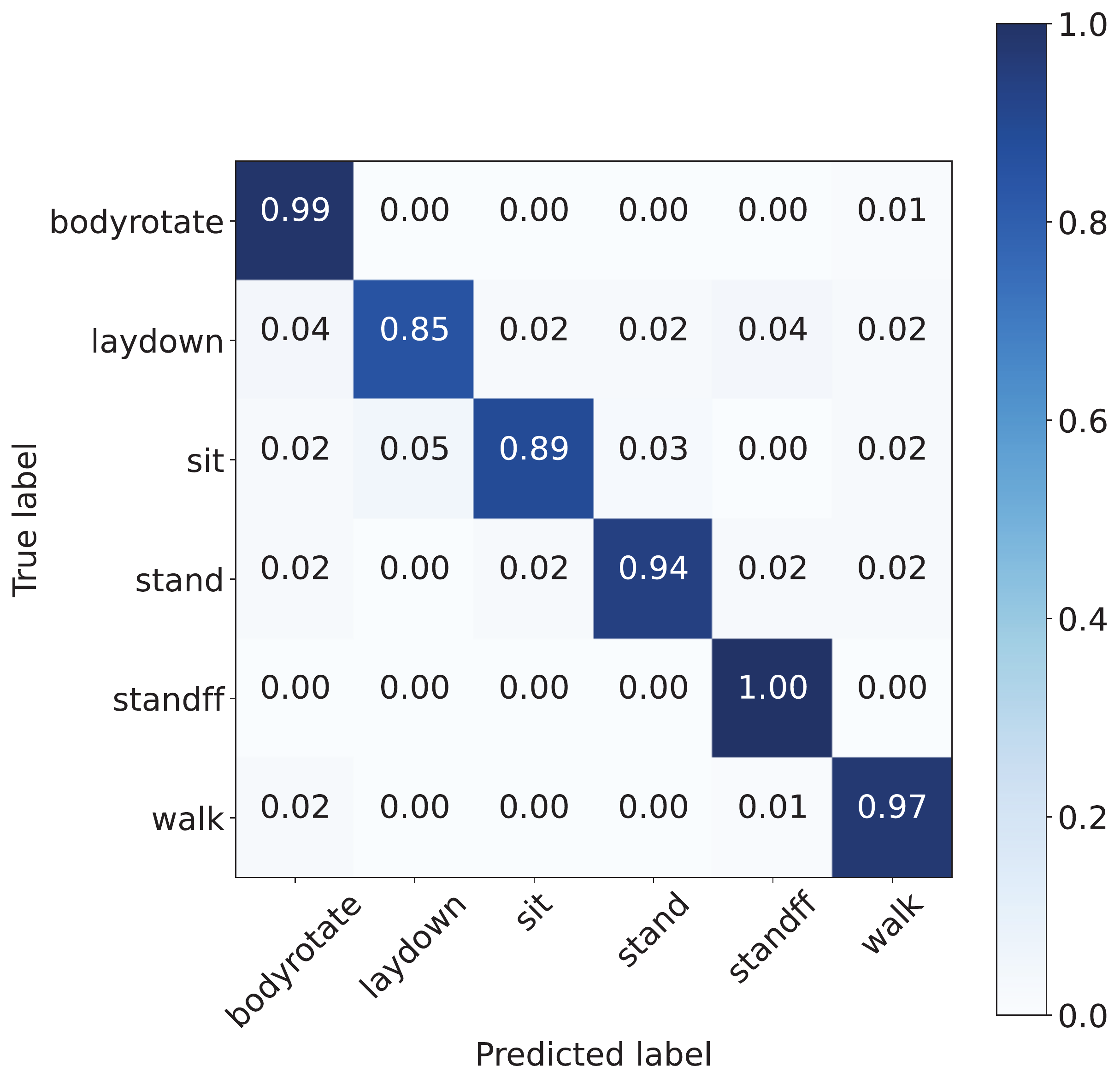}
    \caption{Confusion matrix with the Fusion Transformer model.}
    \label{fig4:fa}
  \end{subfigure}
  \vfill
  \begin{subfigure}[b]{0.8\columnwidth}
    \includegraphics[width=\columnwidth]{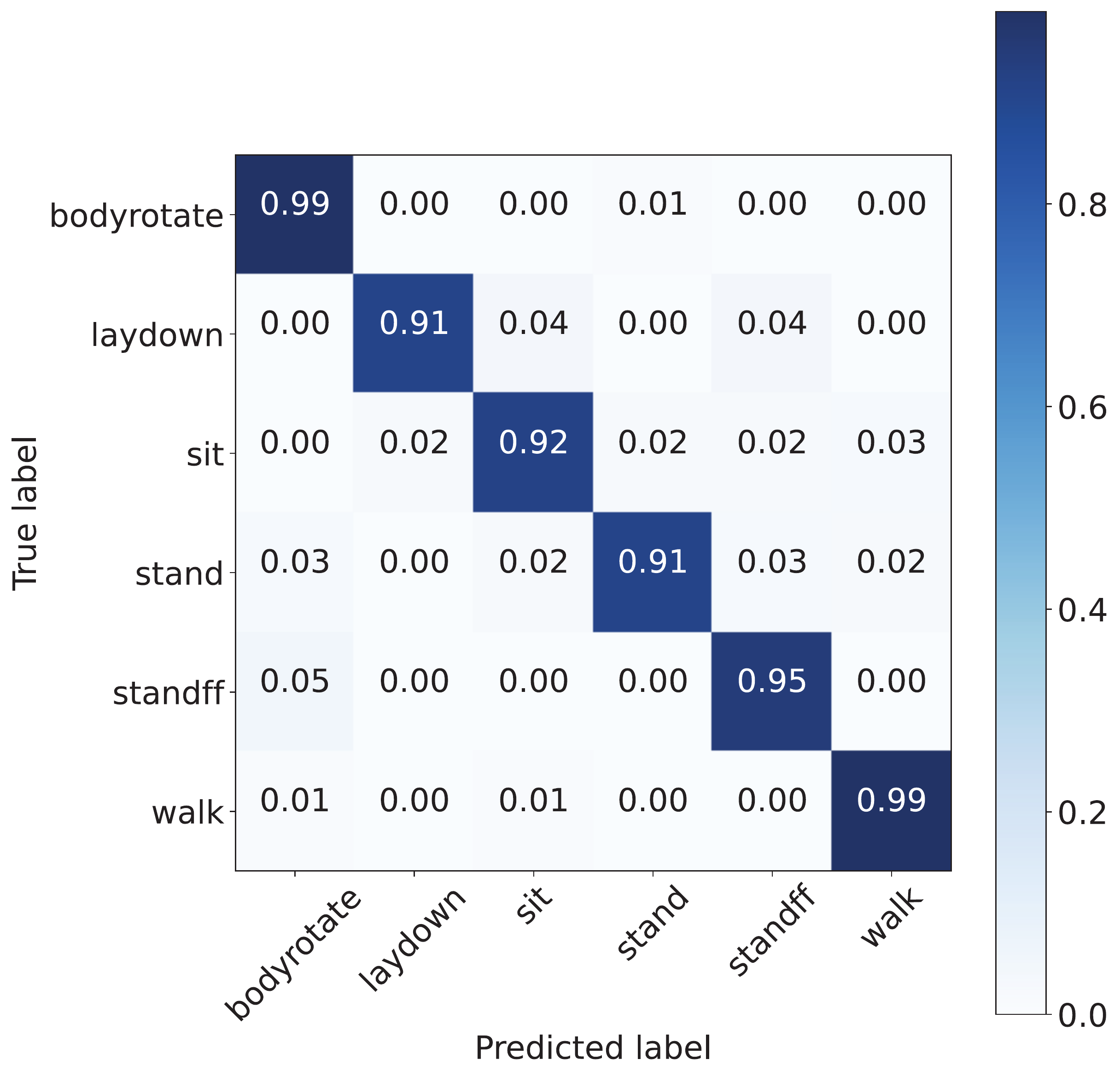}
    \caption{Confusion matrix with the ResNet34 model.}
    \label{fig4:fb}
  \end{subfigure}
  \caption{Visualisation of the confusion matrices of the models with highest scores.}
\label{fig4:f}
\end{figure}

Our experiments showed that when training both our Fusion Transformer and ResNet from scratch, the Fusion Transformer obtained competitive results without any pre-training on a small amount of images. In Table 2, we present the results of SF-ViT, ResNet and Fusion Transformer performance on the validation set when varying the number of image-based features used for training.
With our Fusion Transformer architecture, we obtained our best results when using only PWR and CSI spectrograms, reaching a macro F1-Score of 94.3\%. With ResNet34, we also obtained the best results when using PWR and CSI spectrograms, reaching a F1-score of 94.9\%. The two confusion matrices are shown in Fig. \ref{fig4:f}.
Thus, ResNet34 seems to achieve slightly better performance for HAR. However, the Fusion Transformer can achieve competitive performance with less parameters when trained from scratch, without pre-training. The Fusion Transformer has 11.7M trainable parameters against 12.4M for the ResNet34. One benefit of the Fusion Transformer is that the number of trainable parameters is invariant to the addition of new image-based representations. Unlike the Fusion Transformer, the number of trainable parameters increases with ResNet when doing so.

\section{Towards Self-Supervision}
Transformers outperform many state-of-the-art models when trained on large scale datasets. In this work, we succeeded in achieving competitive results compared to ResNet while training our model from scratch. However, we believe that with self-supervision, the Fusion Transformer can outperform ResNet34 for HAR. Instead of training a model from scratch with weights initialised arbitrarily, the model can be pre-trained via different self-supervised learning (SSL) methods. 

\begin{figure*}[!htp]
    \centering
    \includegraphics[width=0.95\textwidth]{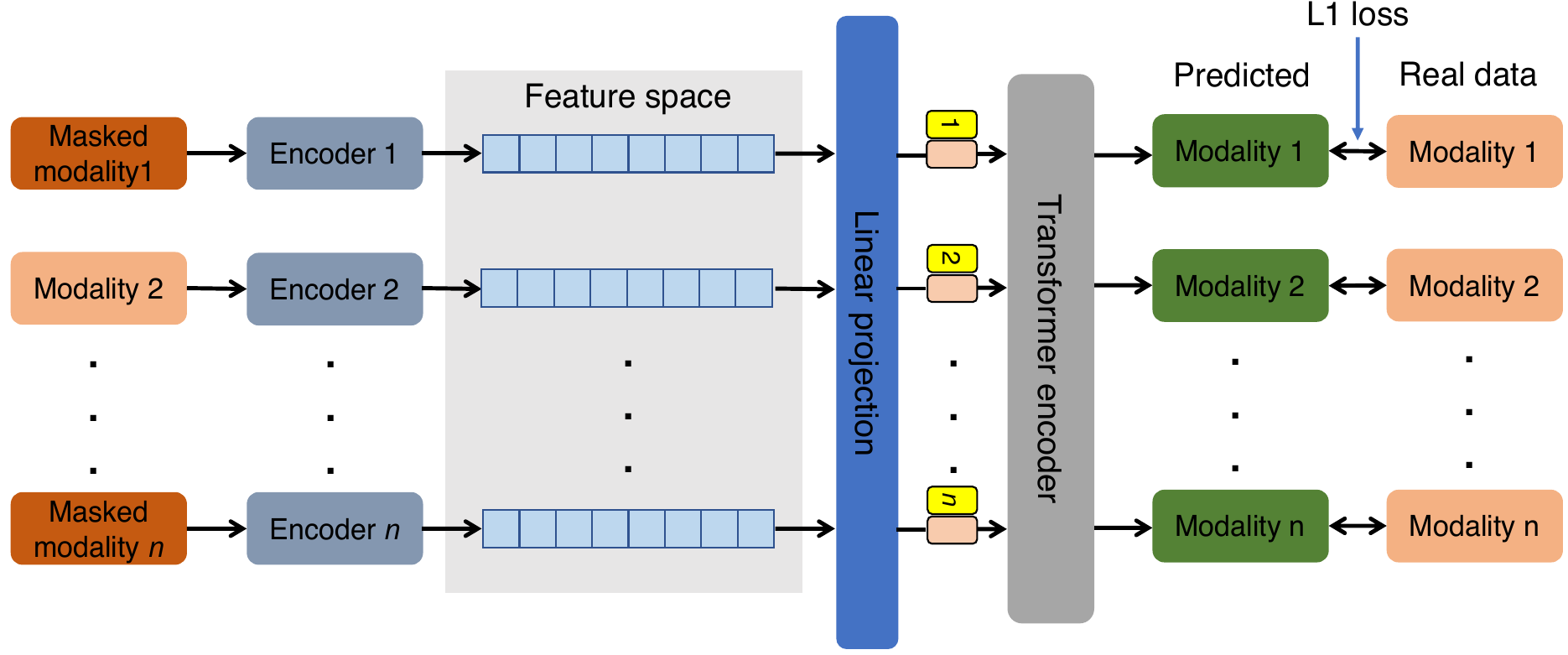}
    \caption{Self-supervised Fusion Transformer method. We randomly mask 60\% of our modalities and we train a model to predict the masked modalities.}
    \label{fig:ssl}
\end{figure*}

We propose a self-supervised method based on image masking as in \cite{simmim}. However, instead of masking some parts of a natural image, in our approach, we mask multiple image features and we pre-train our model to predict the masked image features. The architecture used during the pre-training phase is presented in Fig. \ref{fig:ssl}.
\begin{table*}[!htp]
\centering
\caption{Comparison of pre-trained and supervised Fusion Transformer depending on the size of training set $D$.}
  \begin{tabular}{|c|c|c|c|c|c|c|c|}
    \hline
       & 1 sample  & 2.5\% of & 5\% of & 10\% of & 15\% of & 20\% of & Full \\
       & per class &  $D$ & $D$ & $D$ & $D$ & $D$ & $D$ \\
      \hline
         
      Fusion Transformer (with SSL)    & \textbf{56.3\%} & \textbf{77.0\%} & \textbf{84.5\%} & \textbf{89.7\%} & \textbf{90.4\%} & \textbf{91.2\%} & \textbf{95.9\%} \\ 
    F1-Score &        &        &        &        &        &        &        \\
      \hline 
     Fusion Transformer (no SSL)     & 32.8\% & 60.0\%   & 67\%   & 83.1\% & 84.4\% & 84.4\% & 94.2\% \\ 
      F1-Score &        &        &        &        &        &        &        \\
      \hline
      ResNet34     & 32.6\% & 43.4\%   & 56.9\%   & 62.7\% & 62.2\% & 73.8\% & 94.9\% \\ 
      F1-Score &        &        &        &        &        &        &        \\
      \hline
  \end{tabular}
  \label{tab:sslresults}
\end{table*}
\subsection{Pre-training Phase: Experimental Setup}
We pre-train our model with both PWR and CSI sepctrograms, using all different views and image-based features. We masked 60\% of the image-based features and pre-train our model for 500 epochs. We use an AdamW optimizer and a multi-step learning rate scheduler. The batch size is fixed as 64, the base learning rate as 5e-4, weight decay as 0.05, $\beta_1 = 0.9$, $\beta_2 = 0.999$ and a warm-up for 10 epochs. 
\subsection{Fine-tuning Phase: Experimental Setup and Results}
Next, we fine-tune the pre-trained model in a supervised way. The strength of self-supervised learning is that we can fine-tune the pre-trained model on a smaller training set. This is particularly useful when labelling the data is time consuming and expensive. We train the model on different number of training samples: 1 sample per class, 5\% (10 minutes), 10\% (20 minutes), 15\% (30 minutes), 20\% (40 minutes) of the train set and also the full train set. We fine-tuned our model using the following hyper-parameters settings: an AdamW optimizer, a base learning rate fixed as 1e-3, $\beta_1 = 0.9$, $\beta_2 = 0.999$ and a warm-up for 10 epochs. We added multiple regularisation methods: a weight decay of 0.05 and a stochastic depth \cite{stochdepth} ratio of 0.1. We report the results of our self-supervised method in Table \ref{tab:sslresults}.

ResNet as an architecture is not well-defined for SSL when having multi-modal and multi-sensor data. Existing approaches involve contrastive learning methods \cite{multimodalssl,surveycontrastive}, which require multiple views for each modality, data augmentation techniques and pairs of negative/positive samples. 
A similar work has proposed a self-supervised  contrastive pre-training method for passive Wi-Fi activity recognition\cite{pretrainpwrcsi}.
Two different approaches have been explored in \cite{pretrainpwrcsi}: (1) pre-training the model using two views of CSI data or (2) pre-training the model with one view of CSI data and one view of PWR data. Although we can simply pre-train a model using a contrastive method with two views on many different CNN benchmark models, this framework is not well defined for multi-view and multi-modal pre-training and led to worse results than those presented by our non pre-trained ResNet34. Thus, in the context of multi-sensor fusion with multi-views, we cannot rely on a ResNet architecture.

In this work, we have proposed a simple but yet very effective method for multi-modal and multi-sensor self-supervised learning with a Fusion Transformer which outperforms the results obtained with a non-pre-trained ResNet34 and a non-pre-trained Fusion Transformer, regardless of the training set size. The strength of the Fusion Transformer is that it can be easily pre-trained with multiple views, sensors and modalities, thanks to the transformer architecture.

\section{Conclusion}
We proposed a new architecture for multi-modal, multi-sensor passive Wi-Fi based human activity recognition (HAR). Using signal processing, we extracted 15 image-based features from multiple sensors. With our Fusion Transformer architecture, we first embed each modality via an encoder and then pass it into our transformer network. The Fusion Transformer can fuse multiple image-based features and train a classifier to predict six daily activities performed by six participants. The best results of this model were achieved with PWR and CSI spectrograms, achieving competitive performance with ResNet34, but with less trainable parameters. We next demonstrated that with our proposed self-supervision technique, our pre-trained model outperformed non pre-trained ResNet34, achieving a F1-score of 95.9\%  when fine-tuned on the full training set. Furthermore, it outperformed the other models when fine-tuned with as little as 1\% (2 minutes) of labelled training data with an F1-score of 56.3\%, while the F1-score achieved with 20\% (40 minutes) of training data was 91.2\%.
These results are promising given the need to collect training data for each new indoor environment.

\section*{Acknowledgements}
This work was performed as a part of the OPERA Project, funded by the UK Engineering and Physical Sciences Research Council (EPSRC), Grant EP/R018677/1. 

\bibliographystyle{IEEEtran}
\bibliography{bibliography}
\end{document}